\documentclass[aps,prb,twocolumn,10pt,showpacs,superscriptaddress]{revtex4-1}
\usepackage[dvips]{graphicx}
\usepackage{amssymb}
\usepackage{amsmath}
\usepackage{color}
\usepackage{ulem}
\begin{document}

\title{Measurement of Topological Berry Phase in Highly Disordered Graphene}

\author{K. Bennaceur}
\affiliation{Department of Physics, McGill University, Montr\'eal, Qu\'ebec, H3A 2T8, Canada}

\author{J. Guillemette}
\affiliation{Department of Physics, McGill University, Montr\'eal, Qu\'ebec, H3A 2T8, Canada}
\affiliation{Department of Electrical and Computer Engineering, McGill University, Montr\'eal, Qu\'ebec, H3A 2A7, Canada}

\author{P. L. L\'evesque}
\affiliation{Department of Chemistry, Universit\'e de Montr\'eal, Montr\'eal, Qu\'ebec, H3C 3J7, Canada}

\author{N. Cottenye}
\affiliation{Department of Chemistry, Universit\'e de Montr\'eal, Montr\'eal, Qu\'ebec, H3C 3J7, Canada}

\author{F. Mahvash}
\affiliation{Department of Electrical and Computer Engineering, McGill University, Montr\'eal, Qu\'ebec, H3A 2A7, Canada}
\affiliation{Department of Chemistry, Universit\'e du Qu\'ebec \`a Montr\'eal, Montr\'eal, Qu\'ebec, H3C 3P8, Canada}

\author{N. Hemsworth}
\affiliation{Department of Electrical and Computer Engineering, McGill University, Montr\'eal, Qu\'ebec, H3A 2A7, Canada}

\author{Abhishek Kumar}
\affiliation{NEST, Istituto Nanoscienze-CNR and Scuola Normale Superiore, Piazza San Silvestro 12, 56127 Pisa, Italy}

\author{Y. Murata}
\affiliation{NEST, Istituto Nanoscienze-CNR and Scuola Normale Superiore, Piazza San Silvestro 12, 56127 Pisa, Italy}

\author{S. Heun}
\affiliation{NEST, Istituto Nanoscienze-CNR and Scuola Normale Superiore, Piazza San Silvestro 12, 56127 Pisa, Italy}

\author{M. O. Goerbig}
\affiliation{Laboratoire de Physique des Solides, Univ. Paris-Sud, CNRS UMR 8502, 91405 Orsay, France}

\author{C. Proust}
\affiliation{Laboratoire National des Champs Magn\'etiques Intenses, CNRS, INSA, UJF, UPS, Toulouse, 31400 France}

\author{M. Siaj }
\affiliation{Department of Chemistry, Universit\'e du Qu\'ebec \`a Montr\'eal, Montr\'eal, Qu\'ebec, H3C 3P8, Canada}

\author{R. Martel }
\affiliation{Department of Chemistry, Universit\'e de Montr\'eal, Montr\'eal, Qu\'ebec, H3C 3J7, Canada}

\author{G. Gervais}
\affiliation{Department of Physics, McGill University, Montr\'eal, Qu\'ebec, H3A 2T8, Canada}

\author{T. Szkopek}
\email{thomas.szkopek@mcgill.ca}
\affiliation{Department of Electrical and Computer Engineering, McGill University, Montr\'eal, Qu\'ebec, H3A 2A7, Canada}

\date{\today}

\begin{abstract}
We have observed the quantum Hall effect (QHE) and Shubnikov-de Haas (SdH) oscillations in highly disordered graphene at magnetic fields up to 65 T. Disorder was introduced by hydrogenation of graphene up to a ratio H/C $\approx 0.1\%$. The analysis of SdH oscillations and QHE indicates that the topological part of the Berry phase, proportional to the pseudo-spin winding number, is robust against introduction of disorder by hydrogenation in large scale graphene.
\end{abstract}


\maketitle

\section{Introduction}

Berry phases play an important role in the electronic properties of materials \cite{Berries}. For example, the Berry phase contribution to the closed cyclotron orbits of charge carriers confined to two dimensions in the presence of a perpendicular magnetic field shifts the Landau level (LL) sequence. The Berry phase shift can be observed in the quantum Hall effect (QHE) sequence and the phase of Shubnikov-de Haas (SdH) oscillations. A Berry phase of $\beta = \pi$ has been measured in monolayer graphene \cite{kimqhe, kostyaqhe}, in accordance with the LL sequence for graphene first predicted by McClure \cite{McClure}. The unusual LL sequence of graphene is a consequence of the (pseudo) spinor structure of massless Dirac fermions, where a topological phase shift of $\pi$ is accrued by charge carriers upon $2\pi$ rotation of pseudo-spin in the course of a cyclotron orbit. The $\pi$ Berry phase in monolayer graphene also manifests itself in weak anti-localization that enhances conduction by quantum interference of electrons on closed trajectories, but is readily obscured by the onset of weak localization due to elastic intravalley and intervalley scattering \cite{Antiloc}.\\

The complete Berry phase $\Gamma= \oint_C \mathbf{dk} \cdot i \left<u_{\mathbf{k}} \mid \nabla_{\mathbf{k}} u_{\mathbf{k}} \right>$ of the Bloch wave function $\left.\mid u_{\mathbf{k}} \right>$ on a closed cyclotron orbit is the sum of an energy-dependent non-topological component and a topological component $\beta = \pi W_C$, where $W_C$ is the winding number of the orbit $C$ about valley minimum \cite{Fuchs,Park}. In a semi-classical analysis of LL quantization, the non-topological component of the Berry phase is cancelled by the phase accumulated via orbital diamagnetism \cite{Fuchs}. The anomalous LL sequence for the honeycomb lattice is thus a manifestation of a \textit{topological} Berry phase $\beta = \pi$, that is predicted to persist even in the presence of sub-lattice symmetry breaking and subsequent gap opening in accordance with a full quantum calculation on the honeycomb lattice \cite{Haldane}. In contrast, conventional semiconductors with Schr\"odinger fermions exhibit a total Berry phase $\Gamma = 0$.\\

The question thus arises: how robust is the Berry phase contribution to the LL sequence of graphene in the presence of disorder on the honeycomb lattice? We report here magnetotransport measurements, including SdH oscillations and QHE, of macroscopic hydrogenated graphene monolayers (H/C ratio $\approx 0.1\%$) demonstrating experimentally that the Berry phase remains $\beta = \pi$ in the presence of disorder that is sufficiently strong to impart insulating electron transport behaviour ($dR/dT<0$). Hydrogen adsorbates disrupt the sp$^2$ lattice of graphene through the sp$^3$ distortion necessary to accommodate the C-H bond, and are thus expected to act as neutral point defects. Moreover, a variety of theoretical works show that hydrogenation opens a bandgap in graphene \cite{Duplock04, Sofo07, Boukhvalov08, Lebegue09, Boukhvalov09}, with a recent density functional theory covering a wide range of hydrogen coverage\cite{Rossi15} giving an empirical gap $E_g \approx 3.8 \mathrm{eV} \times (\mathrm{H/C})^{0.6}$. Angle resolved photo-emission spectroscopy (ARPES) and photocurrent measurements give evidence for the onset of insulating electron transport behaviour at H/C$\approx 0.1\%$ \cite{Bostwick09}, with higher hydrogen coverage giving clear evidence of gap formation and the presence of mid-gap states \cite{Grueneis10, Haberer11}. STM imaging \cite{Grueneis12, Goler} has mapped the local density of states in the vicinity of both lone hydrogen adsorbates and hydrogen adsorbate pairs. Previous magnetotransport measurements show that the $\nu = 2$ QHE state can be observed in hydrogenated graphene \cite{QHEGervais}, as well as in disordered graphene grown by sublimation of SiC \cite{Poirier,Andrea}. Thus far, the topological Berry phase in highly disordered graphene has remained unmeasured.

\section{Sample Preparation and Characterization}

Our hydrogenated graphene samples were prepared from pristine monolayer graphene grown by chemical vapor deposition (CVD) on Cu foils \cite{CVD}, and then transferred to oxidized Si substrates for back-gating as previously reported \cite{CVD2}. Electrical contacts (Ti/Au) were deposited by shadow mask technique, thereby minimizing the surface contamination of a lithography process. Both two-point and Hall bar sample geometries were used, as summarized in Table 1. Hydrogenation of graphene on oxidized Si substrates was performed in a UHV chamber with a thermally cracked atomic hydrogen source as previously detailed in Ref.~\onlinecite{QHEGervais}. Varying doses of hydrogen were used, ranging from 4 to 16 minutes at a hydrogen pressure of $\sim 10^{-5}$~Torr. We have observed that the degree of hydrogenation as inferred by Raman spectroscopy and electron transport is coarsely controlled by hydrogen dose, suggesting the possibility of graphene surface contamination or other sources of variability in the hydrogenation process that are not yet understood.

\begin{figure}[!ht]
\includegraphics[width=7cm]{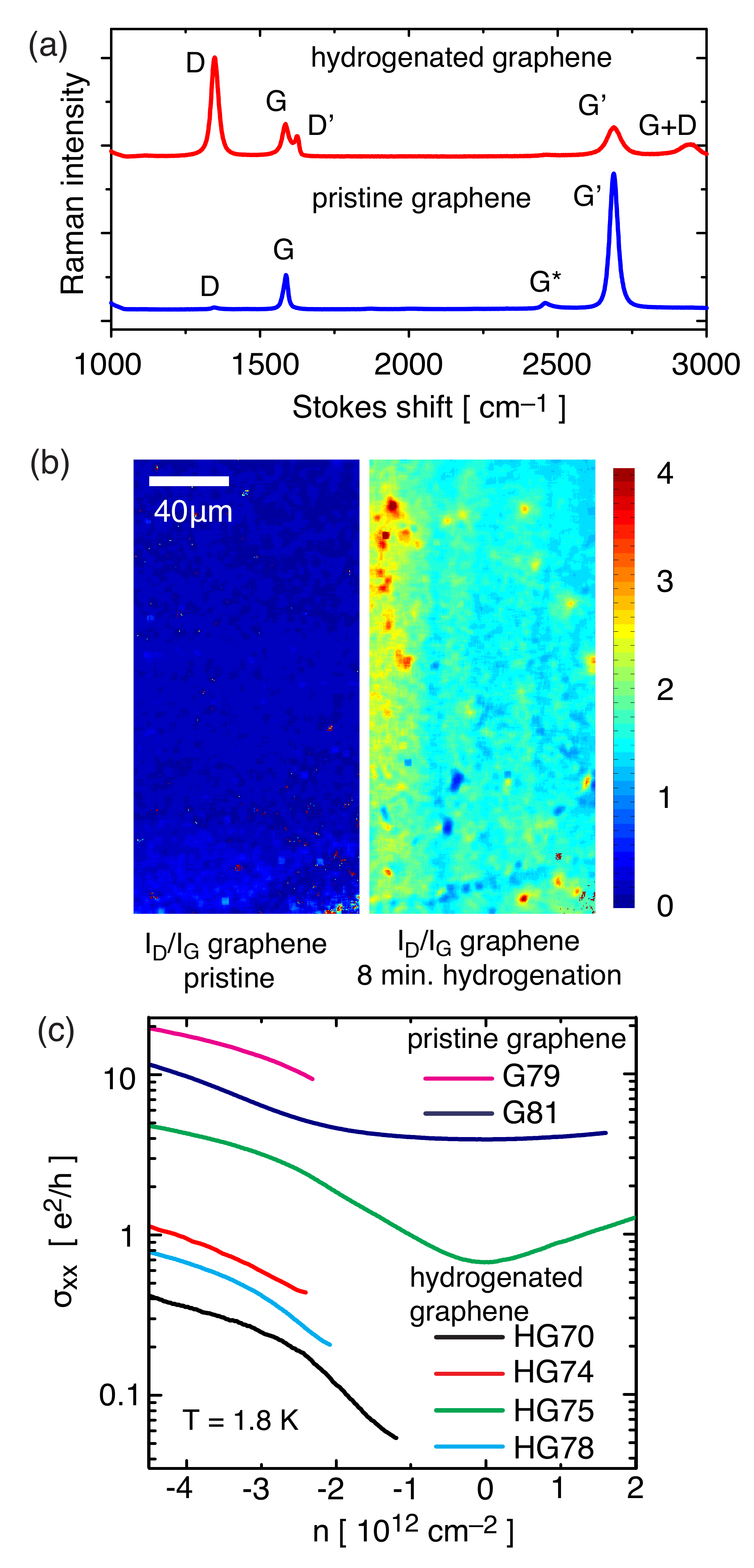}
\caption{\label{Raman}(a) Raman spectrum of  graphene with 4 minutes of hydrogenation exposure (red) and pristine graphene (blue). (b) Raman map of the peak intensity ratio $I_D/I_G$ of the same region of a graphene sheet on SiO$_2$/Si before and after an 8 minute hydrogenation exposure. Macroscopic inhomogeneity is observed, including striations associated with the Cu growth substrate. (c) Conductivity $\sigma_{xx}$ of pristine graphene and hydrogenated graphene versus carrier density $n$ shows a trend of decreasing conductivity upon hydrogenation. The neutrality point, not observable in all samples, was identified from magnetic frequency analysis of Shubikov-de Haas oscillations. }
\end{figure}

The introduction of point defects by hydrogenation was confirmed by Raman spectroscopy performed before and after hydrogenation. Raman point spectra were acquired with an Invia (Renishaw) instrument with an excitation laser line of 25 mW at $\lambda$=514 nm through a 50X objective (plan apochromat, NA=0.75). The laser power on the sample was about $300~\mu\mathrm{W }$ for a laser spot of approximately $1~\mu\mathrm{m}$ in diameter. The spectral region probed was $100-3200~\mathrm{cm}^{-1}$ with a resolution of $\pm 0.5~\mathrm{cm}^{-1}$. An example of a Raman point spectrum is shown in Fig.~\ref{Raman}(a). The $G$, $G^*$, and $G'$ peaks are associated with pristine graphitic carbon, while the $D$, $D'$ and $G+D$ peaks are associated with broken translational symmetry \cite{Jorio11}, which we ascribe to chemisorbed hydrogen whose density can be estimated from the ratio of $D$ peak intensity to $G$ peak intensity, $I_D/I_G$ \cite{Lucchese10}.

Raman spectral images were measured with a RIMA (Photon Etc) hyperspectral imager using tuneable Bragg filters \cite{RIMA}, with an excitation laser line at $\lambda=532~\mathrm{nm}$, a laser power of $40~ \mathrm{\mu W / \mu m^2}$ and an acquisition time of 120 seconds per frame, $1024\times1024$ pixels/frame and spectral resolution of $7~\mathrm{cm}^{-1}$ per frame. The images were acquired with a 50X objective (plan apochromat, NA=0.5) providing a field of view of up to $230\mu\mathrm{m}\times230~\mu\mathrm{m}$ for a spatial resolution of 230~nm per pixel. The spectral regions probed were $1250-1650~\mathrm{cm}^{-1}$ and $2600-2800~\mathrm{cm}^{-1}$. Spectral analysis was performed with D, G, G' Raman Stokes peaks. A third order polynomial background was subtracted followed by fitting to a pseudo-Voigt curve (a combination of Gaussian and Lorentzian broadening) to obtain peak intensity, integrated area, position and full width at half height. The ratio of D peak intensity $I_D$ to G-peak intensity $I_G$ was thus calculated for each pixel in the image. Alignment marks were used to ensure that the same sample area was imaged before and after hydrogenation. A representative Raman map of $I_D/I_G$ measured on the same graphene area before and after hydrogenation is shown in Fig.~\ref{Raman}(b). The striations in $I_D/I_G$ following hydrogenation correlate with the striations in the cold-rolled Cu foil used for graphene CVD growth. With a typical $I_D/I_G \approx 2\pm0.5$ for the samples studied here, we estimate an H/C ratio $\approx 0.07\pm0.02\%$. The extremal Raman intensity ratios $I_D/I_G\approx4$ and $I_D/I_G\approx1$ at inhomogeneities correspond to H/C ratios of $\approx 0.15\%$ and $\approx 0.03\%$, respectively.

\begin{table*}
\caption{Summary of  six hydrogenated graphene samples (HG) and two pristine graphene samples (G) including hydrogenation time (H-time), field effect mobility at a carrier density of $n=-4\times10^{12} \mathrm{cm}^{-2}$ and temperature $T=1.8~\mathrm{K}$. The back-gate voltage of the neutrality point is estimated from the maximum two-point resistance at $T=1.8~\mathrm{K}$. Samples HG78 and G79 were measured twice with two different doping levels as controlled by an applied gate voltage during cool-down from room temperature to fix a charge density on the oxide substrate. The maximum magnetic field applied in our experiments and the sample geometry are reported: two-point configuration (2pt) or Hall bar configuration (HB). The Berry phase, $\beta / 2\pi$, is also reported. The QHE sequence was observed in G79. SdH oscillation phase was directly measured in samples HGTO1, HGTO2, HG75, and HG78. Insufficient SdH oscillations were observed in G81, HG70 and HG74 to extract Berry phase. Further details of phase extraction are given in the text. }
\begin{ruledtabular}
\begin{tabular}{ c c c c c c c }

    Sample & H-time [min] & Geometry & Mobility $\mu_\mathrm{FE}$ [cm$^{2}$/V$\cdot$s] & Neutrality [V] & Max. field [T] & Berry phase $\beta/2\pi$ \\ \hline
    HGTO1 & 8 & 2pt & 100 & 30 V & 65 T &  $-0.73\pm0.03$ \\ \hline
    HGTO2 & 5 & 2pt & 200 & 28 V & 55 T & $-0.64\pm0.08$ \\ \hline
    HG70 & 4 & HB & 21 & 7 V & 35 T & - \\ \hline
    HG74 & 6 & HB & 55 & 80 V & 35 T & - \\ \hline
    HG75 & 7 & HB & 260 & 30 V & 35 T & $-0.49\pm0.11$ \\ \hline
    HG78 & 10 & HB & 20-35 & 25-65 V & 35 T & $-0.53\pm0.14$ \\ \hline
    G79 & 0 & HB & 750-1050 & 10-85 V & 35 T & -1/2 \\ \hline
    G81 & 0 & HB & 550 & 24 V & 35 T & - \\ 

    \end{tabular}
\end{ruledtabular}
\end{table*}

Scanning tunnelling spectroscopy (STS) and microscopy (STM) were performed on pristine and hydrogenated graphene using an RHK Technology STM in an ultra-high vacuum (UHV) chamber with a base pressure of $1\times 10^{-10}$ mbar at 300K. The STM was equipped with a Tectra hydrogen source that produces atomic hydrogen by thermal cracking. After introducing CVD graphene samples into the UHV chamber, they were annealed at 673 K for 2 hours in order to remove water and other adsorbates. At this temperature, it was observed that hydrogen does not desorb from similar samples of graphene on silicon carbide \cite{Goler}. The temperature was measured by a type K thermocouple at the position of the sample and cross-calibrated by an optical pyrometer. Despite the cleaning process, atomic resolution STM images of pristine graphene were not attained, most likely due to residual polymer from the PMMA handle used in the transfer process of large area graphene. Comparison of the atomic hydrogen exposure dose versus the observed $I_D/I_G$ Raman peak ratio indicates a low atomic hydrogen sticking coefficient of $\approx 10^{-4}$, consistent with a polymer residue layer that suppresses atomic hydrogen adsorption to the graphene lattice. For comparison, a sticking coefficient of $\approx 1$ is observed for hydrogen incident on graphene grown directly on silicon carbide without a polymer transfer process\cite{Goler}. A gap was not observed in STS of graphene hydrogenated prior to introduction to the STM UHV chamber, and graphene hydrogenated within the STM UHV chamber with the same hydrogen dose. At an H/C coverage of $\approx 0.07\pm0.02\%$, the gap is theoretically predicted\cite{Rossi15} to be $50\pm10$ meV, below the energy resolution limit of room temperature STS \cite{Morgenstern}.

The measured sheet conductivity $\sigma_{xx}$ at $T= 1.8~\mathrm{K}$ is plotted as a function of carrier density in Fig.~\ref{Raman}(c). The carrier density range over which conductivity could be measured was limited by both accessible gate voltage range, and the effect of shunt capacitance to the back gate for low graphene conductivities. The neutrality point was identified from magnetic frequency analysis of Shubnikov-de Haas oscillations, described further below. A general trend of reduced conductivity upon hydrogenation is observed. All samples, pristine and hydrogenated, displayed insulating behaviour, $dR/dT<0$, upon cooling in zero magnetic field. For all samples reported in this work, several relevant parameters are summarized in Table 1. Pristine graphene samples typically exhibited a conductivity $\sigma_{xx} \sim 10 e^2/h$ at $T= 1.8~ \mathrm{K}$ and at a hole density $\sim3\times10^{12}\mathrm{cm}^{-2}$ tuned by application of a back gate potential, with a corresponding Ioffe-Regel disorder parameter $(k_F \lambda)^{-1} \simeq (2e^2/h)/\sigma_{xx} \sim 0.2$ , where $k_F$ is the Fermi wave-vector and $\lambda$ the carrier mean free path. In contrast, hydrogenated graphene samples typically exhibited a conductivity $\sigma_{xx} \sim 0.5 e^2/h$ at $T= 1.8~\mathrm{K}$ and at a hole density $\sim3\times10^{12}\mathrm{cm}^{-2}$, with a corresponding Ioffe-Regel disorder parameter $(k_F \lambda)^{-1} > 1$. The two--point resistance $R_{2pt}$ at charge neutrality was generally found to exhibit steeper temperature dependence $dR/dT$ for samples with larger Ioffe-Regel disorder parameter. We have previously observed that hydrogenated samples with disorder parameter well beyond the Ioffe-Regel limit for metallic conduction are still capable of supporting a QHE state \cite{QHEGervais}.

\section{Magneto-transport}

The resistance of six hydrogenated graphene samples and two pristine graphene samples were measured at low temperature, $0.3\mathrm{K}-1.5\mathrm{K}$, and high magnetic field. Quasi-dc magnetic fields were applied by either a 35 T resistive magnet or a 45 T hybrid resistive/superconducting magnet at NHMFL Tallahassee. Pulsed magnetic fields were applied at LNCMI Toulouse up to 55 T (28 mm bore) or 65 T (13 mm bore) \cite{magnet}. The two-point resistance $R_{2pt}$ was measured for all samples. The longitudinal resistance $R_{xx}$  and the Hall resistance $R_{xy}$ were measured on Hall bars. Sample resistance was measured by standard lock-in technique with quasi-dc magnetic fields, while high-speed baseband measurements were used in pulsed magnetic fields. The carrier density was tuned with a back-gate voltage.

\begin{figure}[!ht]
\includegraphics[width=7cm]{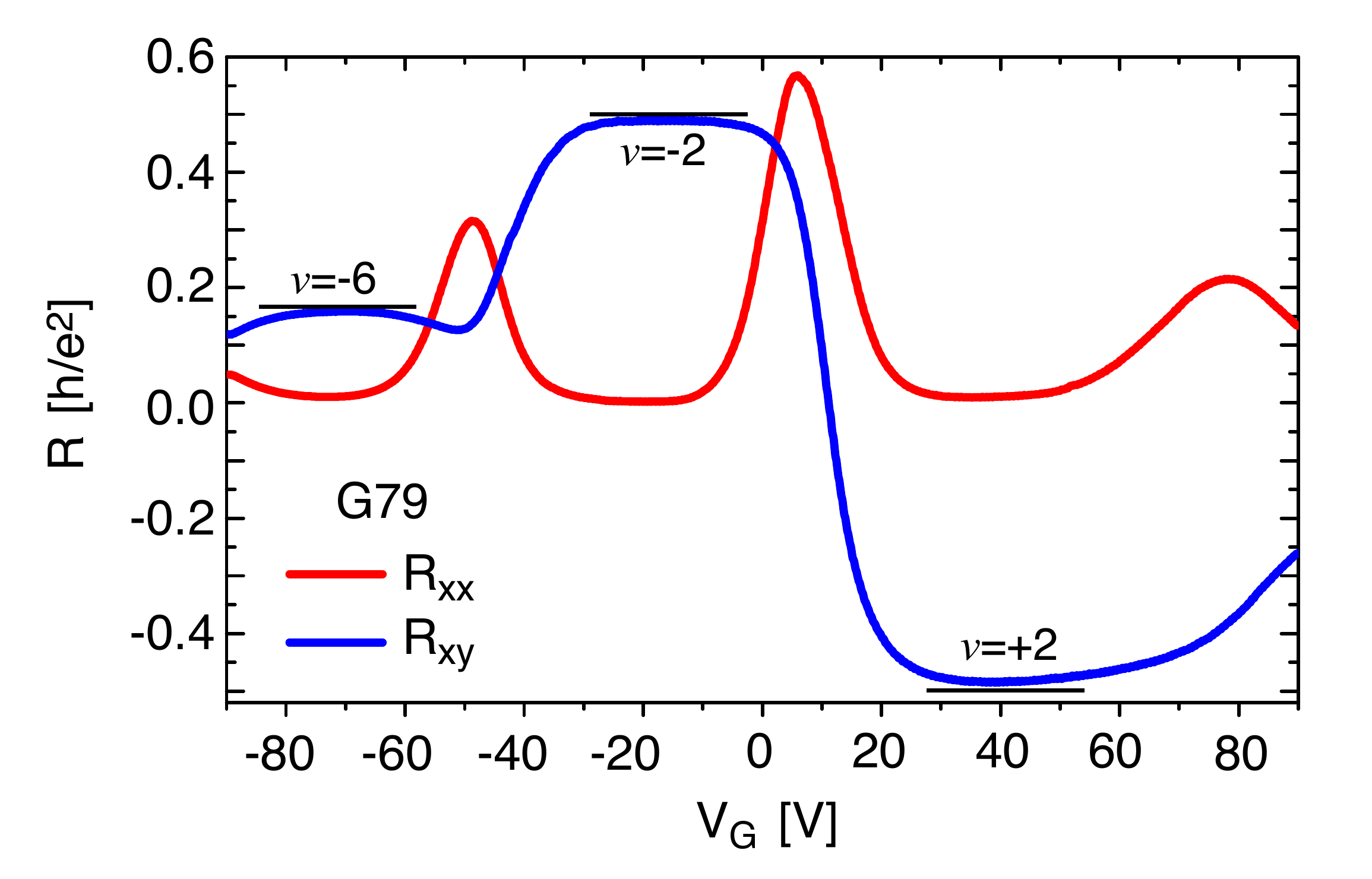}
\caption{\label{cleanQHE}The measured longitudinal resistance $R_{xx}$ and Hall resistance $R_{xy}$ of a large area CVD graphene sample G79 versus back gate voltage $V_G$ at temperature $T=0.3$ K and constant magnetic field $B = 35$ T. The observed QHE filling factor sequence $\nu = -6, -2, +2$ corresponds to the anomalous Hall sequence for massless Dirac fermions in monolayer graphene.}
\end{figure}

The longitudinal and Hall resistance of large area graphene sample G79 is shown in Fig.~\ref{cleanQHE} at low temperature $T=0.3$ K and constant magnetic field $B = 35$ T. The longitudinal resistance $R_{xx}$ is minimized while the Hall resistance $R_{xy}$ approaches quantized values $h/\nu e^2$, identifying the QHE filling factors $\nu = -6, -2, +2$.  The filling factor sequence corresponds to the LL sequence of massless Dirac fermions with topological Berry phase $\beta/2 \pi = \pm1/2$. Prior to hydrogenation, CVD-grown millimetre scale graphene exhibits the same LL sequence as that observed in pristine exfoliated graphene \cite{kimqhe, kostyaqhe}.\\

\begin{figure}[!ht]
\includegraphics[width=7cm]{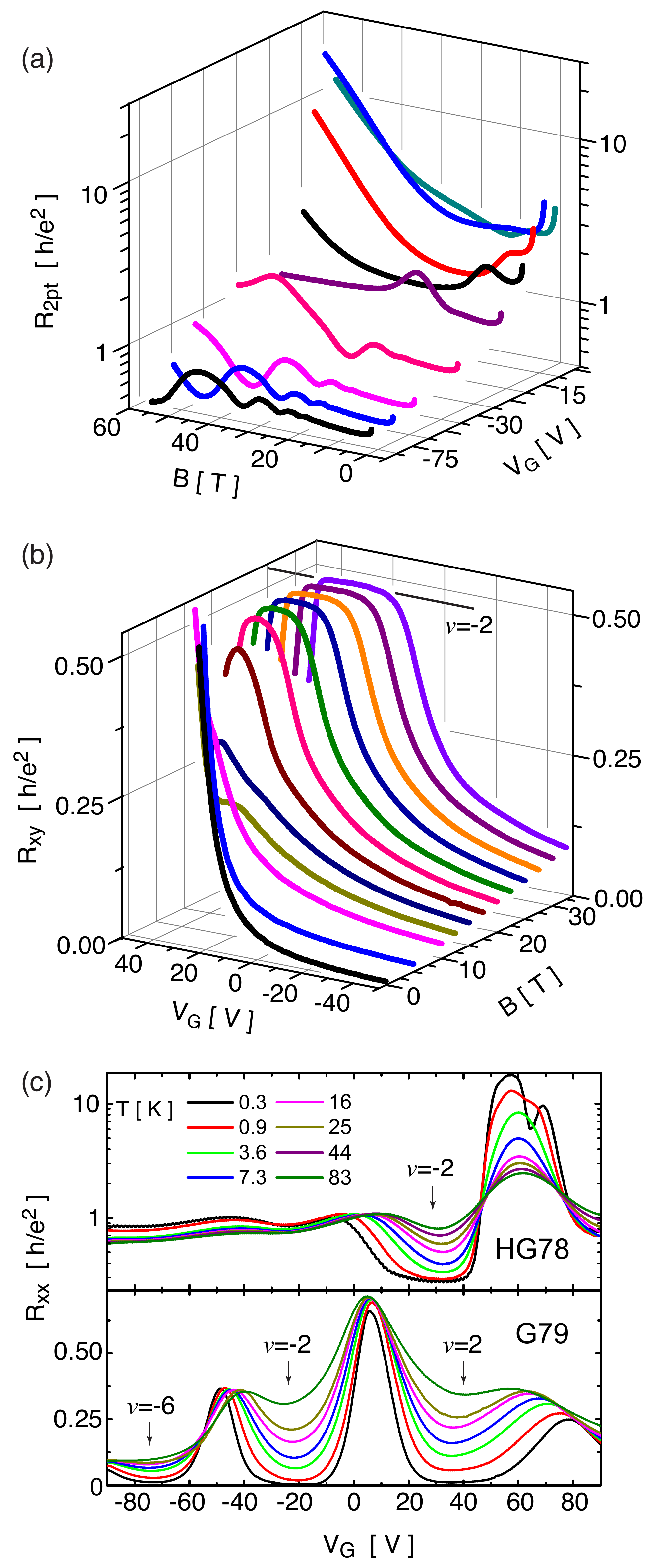}
\caption{\label{QHE}(a) The measured $R_{2pt}$ of hydrogenated graphene sample HGTO2 on a logarithmic scale versus applied magnetic field $B$ and back-gate voltage $V_G$, at a sample temperature $T = 1.5~\mathrm{K}$. SdH oscillations and a strong positive magnetoresistance near neutrality are observed. (b) The measured $R_{xy}$ of hydrogenated graphene sample HG78 on a logarithmic scale versus applied magnetic field $B$ and back-gate voltage $V_G$, at a sample temperature $T = 0.3~\mathrm{K}$. A $\nu=-2$ plateau at $R_{xy}=12550~\Omega$ is observed above a field of $20~\mathrm{T}$.  (c) The measured $R_{xx}$ of hydrogenated graphene sample HG78 (log-scale) and pristine graphene sample G79 (linear scale) versus applied gate voltage, over a temperature range $T =~0.3~\mathrm{K}-83~\mathrm{K}$ at a constant field of $B = 35 \mathrm{T}$. The minima in $R_{xx}$ are labelled with corresponding filling factors.}
\end{figure}

Representative magneto-transport measurements of hydrogenated graphene samples are shown in Fig.~\ref{QHE}. The measured $R_{2pt}$ of hydrogenated graphene sample HGTO2 versus magnetic field at fixed gate voltages is shown in Fig.~\ref{QHE}(a). The magnetoresistance shows weak localization at small fields,  SdH oscillations, and a strong positive magnetoresistance at charge neutrality. The measured Hall resistance $R_{xy}$ of sample HG78 versus gate voltage at fixed magnetic field is shown in Fig.~\ref{QHE}(b). At applied magnetic fields exceeding $20~\mathrm{T}$, a plateau emerges at $R_{xy} = 12550~\Omega$, within $3\%$ of $h/2e^2$ for a $\nu = -2$ QHE state. The $\nu=-2$ QHE state has been previously observed in hydrogenated graphene, and all samples studied here exhibited evidence of a $\nu = -2$ QHE state.\\

The temperature dependence of $R_{xx}$ versus gate voltage at $B = 35~\mathrm{T}$ is shown in Fig.~\ref{QHE}(c) for a pristine graphene sample (G79) and a hydrogenated graphene sample (HG78). The minima of $R_{xx}$ at $\nu=-2$ fit well with a Mott variable range hopping law $R_{xx}^{min} \propto \sigma_{xx}^{min} \propto \exp(-(T_{0}/T)^{1/3})$ for both hydrogenated and pristine graphene, as has been observed in pristine exfoliated graphene \cite{Keyan}. Due to large Landau level separation in graphene ($\simeq$ 2500~K for $\nu=2$ at 35~T), we expect to observe intra-LL transport described theoretically by a VRH model over the temperature range probed in our experiment. The $R_{xx}$ at neutrality shows weak temperature dependence for pristine graphene G79. In contrast, $R_{xx}$ of hydrogenated graphene HG78 grows over one order of magnitude at neutrality over the same temperature interval from $T=83~\mathrm{K} - 0.3\mathrm{K}$. Similar behaviour has been observed in pristine graphene and interpreted as a phase transition to a strongly insulating $N=0$ LL \cite{Zhang06, Checkelsky08, Zhang10} bounded by critical points of temperature independent $R_{xx}$. The mechanism by which hydrogenation induces a strongly insulating $N=0$ LL is unknown, and we speculate that local sub-lattice symmetry breaking in the immediate vicinity of isolated hydrogen adsorbates may play a role. Hydrogenated graphene samples exhibited electron-hole asymmetry in the form of lower electron conduction as compared to hole conduction, as predicted theoretically for both zero-field \cite{AsymEH} and high-field \cite{Roche} transport. The origin of the dip in $R_{xx}$ of sample HG78 near neutrality at $V_G = 65 \mathrm{V}$ and $T=0.3 \mathrm{K}$ is unclear.

\begin{figure}[ht]
\includegraphics[width=7cm]{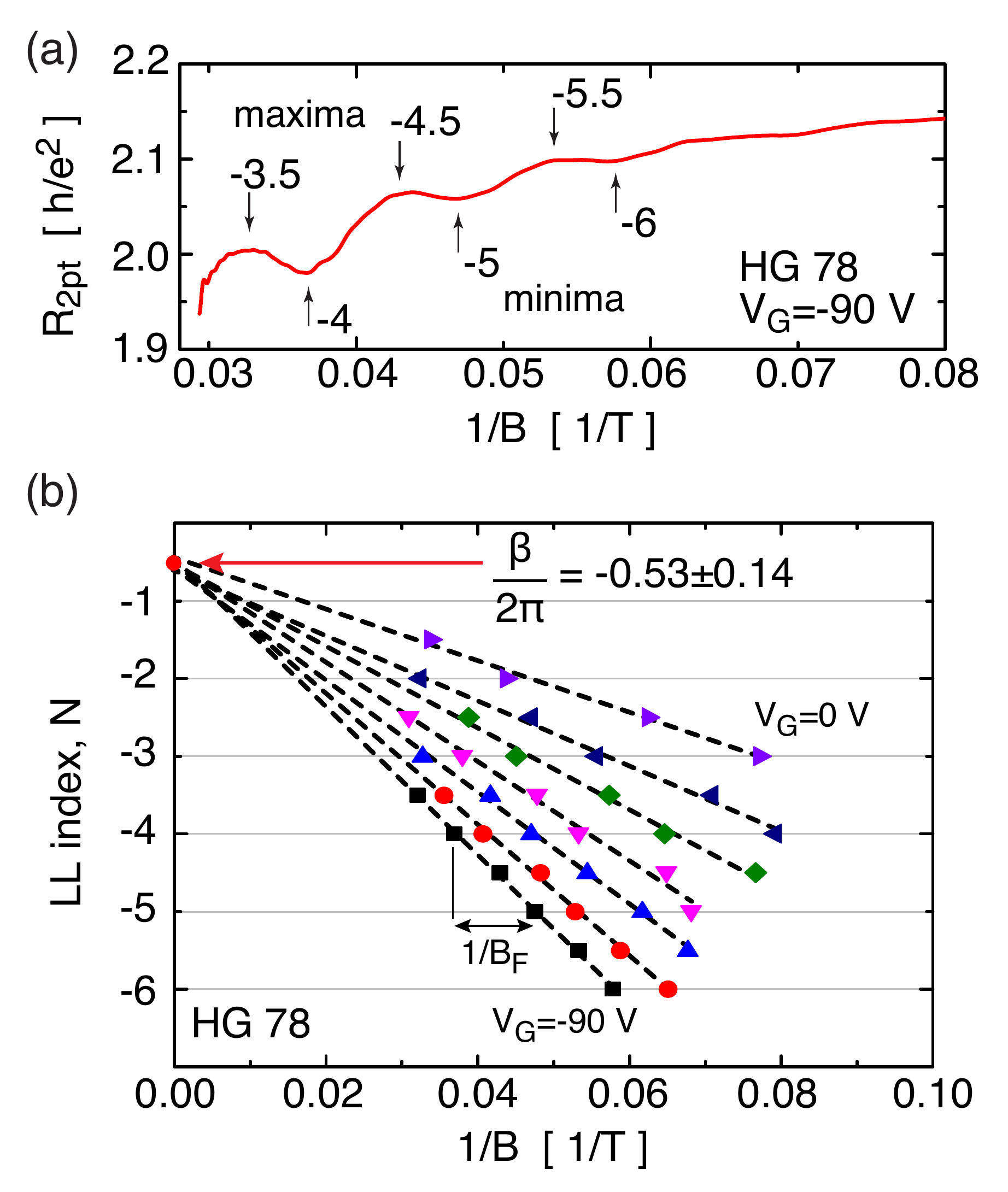}
\caption{\label{Bf4}(a) The measured $R_{2pt}$ versus $1/B$ of HG78 at a gate voltage $V_G = -90~\mathrm{V}$. The local minima and maxima of $R_{2pt}$ versus $1/B$ are identified. (b) A Landau fan diagram of LL index versus $1/B$ for sample HG78 at different gate voltages. The SdH frequency $B_F$ at each gate voltage is extracted from the slope of the LL fan (linear fit indicated by dashed lines). The intercept $\beta / 2 \pi = -0.53\pm 0.14$ is in good agreement with $\beta = \pi$.}
\end{figure}

\section{Landau Level Analysis}

We now turn our attention to the LL sequence inferred from SdH oscillations, generally expected to be of the Lifshitz-Kosevich form $\Delta R_{xx}=R(B,T) \cos[2\pi(B_{F}/B+1/2)+\beta]$ where $B_{F}$ is the frequency of the oscillations, $\beta$ is the (topological) Berry phase, and $R(B,T)$ is an envelope dependent upon both temperature and field. In the limit where $R_{xy} \ll R_{xx}$, we find the SdH oscillations $\Delta R_{2pt} \propto \Delta R_{xx}$.

To determine the Berry phase, we constructed Landau fan diagrams, as shown in Fig.~\ref{Bf4}(b) for sample HG78, in which the index $N$ corresponding to the $N^{th}$ minimum in $R_{2pt}$ is plotted versus the reciprocal field $1/B$ at which the minimum occurs for a range of back-gate voltages $V_G$. Also plotted are indices $N+1/2$ for the $N^{th}$ maxima in $R_{2pt}$ versus the reciprocal field $1/B$ at which the maxima occur. At each back gate voltage, the slope of $N$ versus $1/B$ (or $N+1/2$ versus $1/B$ ) gives the frequency $\delta N/\delta(1/B) = B_F$. The intercept at $1/B = 0$, which we determine by extending a linear fit at each gate voltage, corresponds to the Berry phase shift $\beta /2\pi$.\cite{kimqhe} For the sample HG78, the intercept gives a Berry phase $\beta = -\pi\cdot(1.06\pm0.28)$ in good agreement with $\beta=\pi$ modulo $2\pi$. We estimate Berry phase error from the standard error for the intercept in a linear regression fit of $N$ versus $1/B$.  For comparison, the Berry phase experimentally extracted by Landau fan diagram intercept for pristine graphene in the hole doped regime has been reported to be $\beta \approx -\pi \cdot 1.2$. \cite{kimqhe} The normalized topological Berry phase can be identified with Diracness, a metric quantifying SdH phase between that of Schr\"odinger and massive Dirac fermions \cite{Diracness}, as $\delta = |\beta|/\pi$. One might have expected, from tight-binding calculations, a slight deviation from pure Dirac fermions, $\delta\simeq 1-0.2\Delta/t$, in terms of the energy gap $\Delta$ at K,K' and the nearest-neighbour hopping parameter $t\simeq 3$ eV, due to a contribution from second-nearest-neighbour hopping. However, this deviation is below our experimental resolution.

\begin{figure}[ht]
\includegraphics[width=6.5cm]{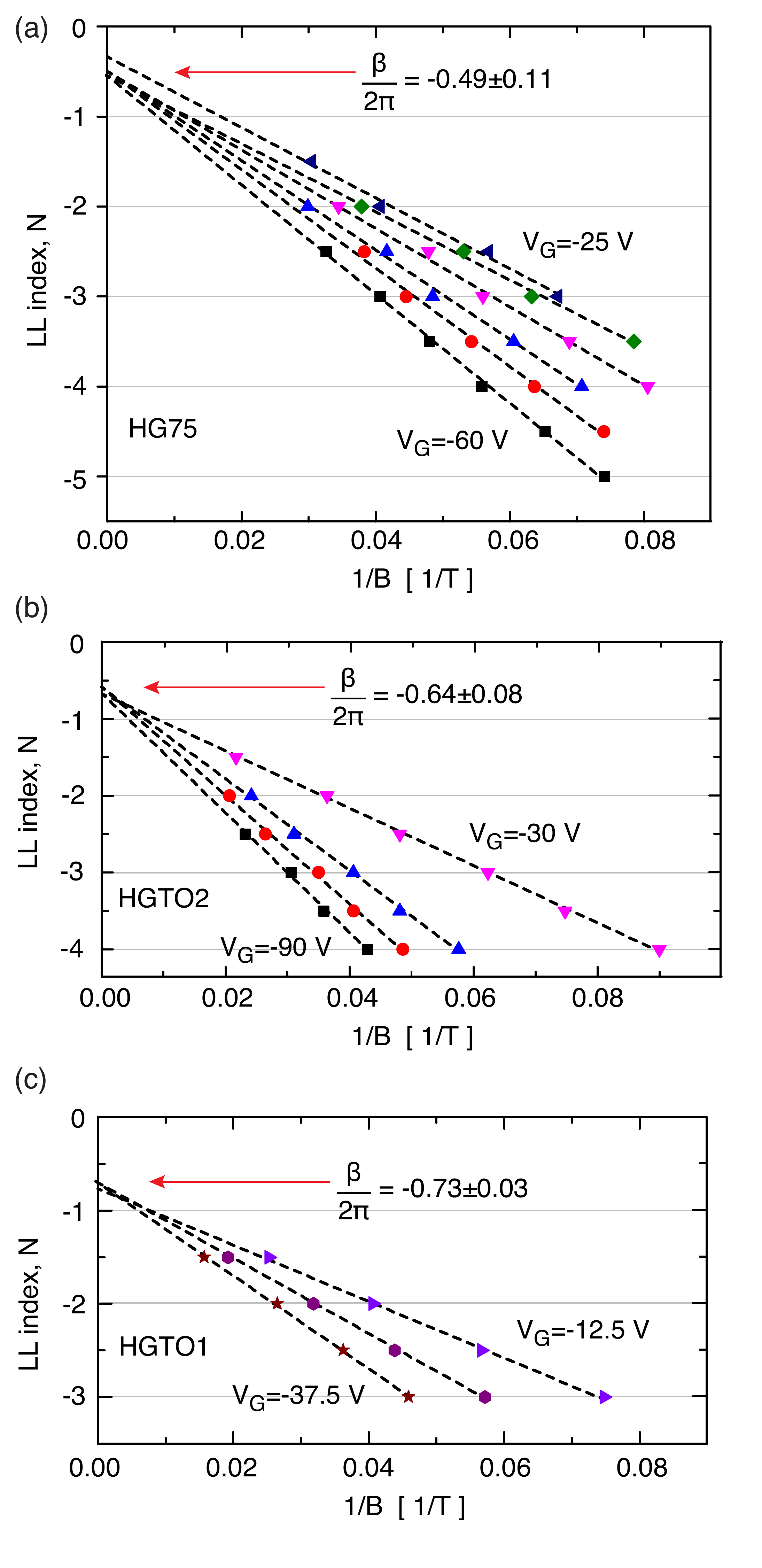}
\caption{\label{fig5}The Landau fan diagrams of samples (a) HG75 at $T=0.3~\mathrm{K}$, (b) HGTO2 at $T=1.5~\mathrm{K}$ and (c) HGTO1 at $T=1.5~\mathrm{K}$. The topological Berry phase $\beta$ is extracted for each sample from the intercept at $1/B = 0$ of the best-fit lines to at least two LL's. }
\end{figure}

Landau fan diagrams constructed from the two-point resistance $R_{2pt}$ of samples HG75, HGTO2 and HGTO1 are shown in Fig.~5. The amplitude of SdH oscillations were observed to be larger in $R_{2pt}$ than in $R_{xx}$, and were thus analyzed for improved signal to noise ratio at the expense of a systematic error introduced by the addition of Hall resistance $R_{xy}$ component in $R_{2pt}$. The accuracy of the extraction of topological Berry phase, $\beta$, from the intercept at $1/B = 0$ improves with increasing number of SdH oscillations, and we therefore only used measurements performed at gate voltages (carrier densities) that resulted in at least two minima and two maxima in $R_{2pt}$. The indicated error in $\beta/2\pi$ for each sample is the standard deviation of the mean over different carrier densities. The normalized Berry phase $\beta / 2 \pi$ measured from Landau diagram intercepts for samples HG75, HGTO2 and HGTO1 are $-0.49\pm0.11$, $-0.64\pm0.08$ and $-0.73\pm0.03$, respectively. The lowest mobility samples, HG70 and HG74, did not give a sufficient number of SdH oscillations to determine Berry phase. As the disorder in our samples is high, local density fluctuations are to be expected. Resistivity measurement gives an average over the entire sample, including local density fluctuations. Consequently, the observed SdH oscillations contain components with slightly different magnetic frequencies that introduce uncertainty in Berry phase extraction. This is especially the case in samples HGTO1 and HGTO2 for which fewer resistivity maxima and minima could be identified. Similarly, analysis of $R_{2pt}$ is anticipated to introduce systematic error.

\begin{figure}[ht]
\includegraphics[width=7cm]{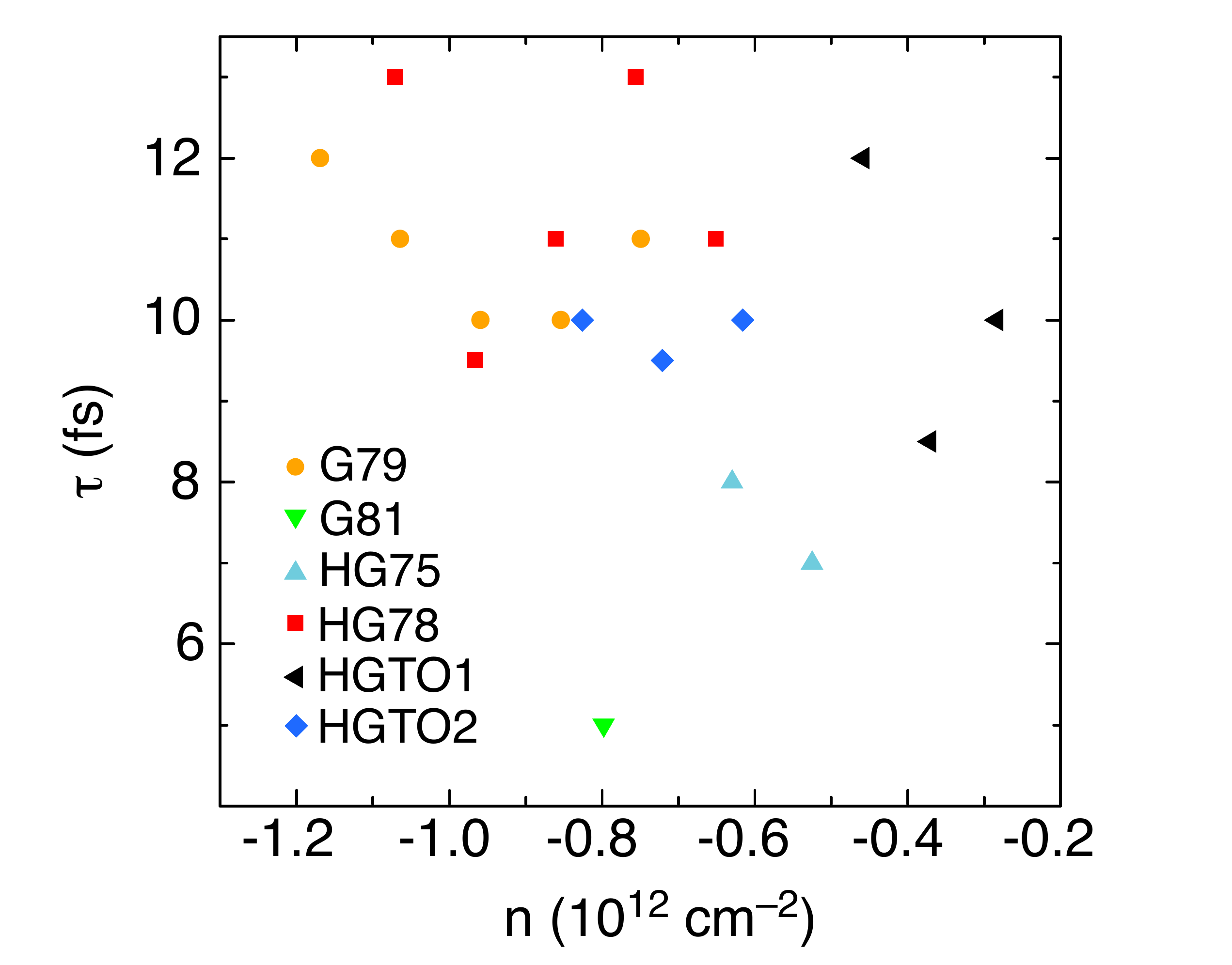}
\caption{\label{Bf6}The scattering time $\tau$ extracted by a Lifshitz-Kosevich best-fit to SdH oscillations with a Dingle damping factor $R(B,T)=R_0 \cdot \exp( -\pi \hbar k_F / e B v_F \tau )$, plotted versus carrier density $n$, for six graphene samples.}
\end{figure}

We next consider the damping of the observed SdH oscillations caused by disorder--induced LL broadening. We approximate the envelope $R(B,T)$ with a simple Dingle damping factor accounting for the linear graphene dispersion given by $R(B,T)=R_0 \cdot \exp( -\pi \hbar k_F / e B v_F \tau )$ where $v_F=1.0\times10^6$m/s is the graphene Fermi velocity and $k_F$ is the Fermi wave-vector. The Fermi wave-vector is in turn inferred from magnetic frequency via the Onsager relation $B_F = \Phi_0 k_F^2 /4\pi$, with $\Phi_0 = h/e$ the flux quantum. The disorder--induced effective scattering time $\tau$ was determined by a best fit of the measured resistance oscillations to $R(B,T) \cos[2\pi(B_{F}/B+1/2)+\beta]$. The inferred scattering times are plotted in Fig.~\ref{Bf6} for six samples versus charge carrier density $n$. The observed scattering times $\tau=$ 5-13 fs correspond to Dingle temperatures $T_D = \hbar/2\pi k_B \tau=$ 90-240 K that characterize the observed LL broadening. Surprisingly, the variation in LL broadening amongst samples is small, varying by less than a factor of three, and it is the pristine graphene sample G81 that suffers the greatest LL broadening, indicative of a common inhomogeneous broadening mechanism. The charge density fluctuation of electron-hole puddles common to graphene on oxide substrates is the most likely source of the LL broadening observed in our work. We note that in our previous work with more heavily hydrogenated graphene samples \cite{QHEGervais}, a direct transition from the insulating state to the $\nu=-2$ QHE state was observed without the observation of SdH oscillations at lower magnetic fields.\\

\begin{figure}[ht]
\includegraphics[width=7cm]{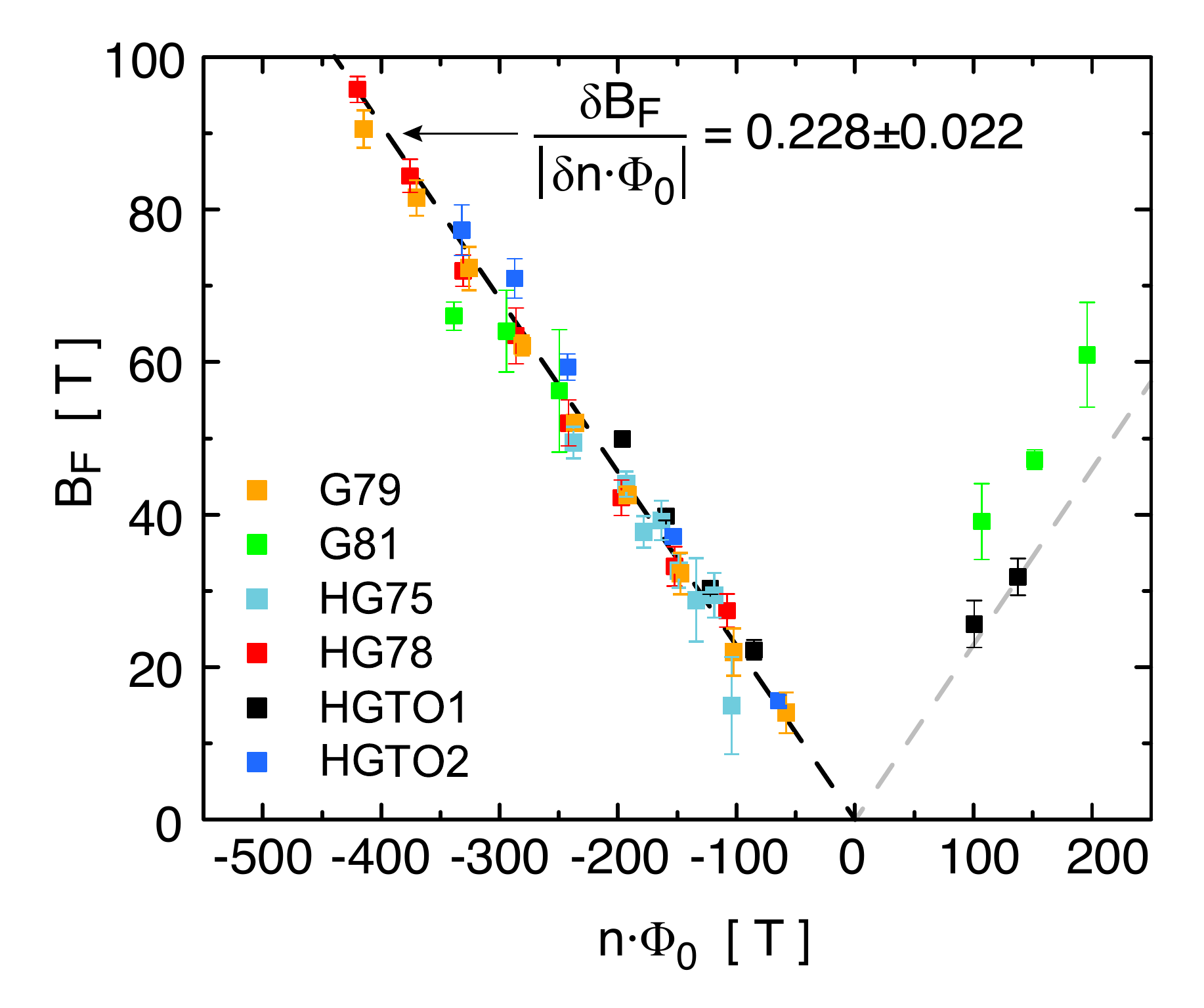}
\caption{\label{tau}The extracted SdH frequency $B_F$ versus carrier density scaled by the flux quantum $n \Phi_0$ for six graphene samples. The best fit slope (black dashed line) on the hole side for hydrogenated graphene samples is $\delta B_F/ |\delta n \Phi_0|=0.228\pm0.022$, in agreement with 4-fold LL degeneracy. Extension of the best fit slope to the electron side is shown with a grey dashed line.  }
\end{figure}

Finally, we analyzed the LL sequence in further detail. The degeneracy $g$ of the LLs can be experimentally determined from the relation $B_F = n\Phi_0/g$, where $n$ is the charge carrier density. The magnetic frequency $B_F$ versus $n \Phi_0$ is shown in Fig.~\ref{tau} for a variety of graphene samples. For each sample, the charge neutrality point ($n=0$) and slope $\delta B_F / | \delta n \Phi_0| = 1/g$ was determined by a linear fit of $B_F$ versus $\Delta n \Phi_0 = (h/e^2) \cdot C V_G$ where $\Delta n$ is the carrier density induced by field effect through the back-gate capacitance $C=11.5~\mathrm{nF/cm^2}$. For the four hydrogenated graphene samples with at least two SdH oscillations observed, the slope $\delta B_F / |\delta n \Phi_0| = 0.228 \pm 0.022$ on the hole side, corresponding to a LL degeneracy $g=4.4 \pm 0.4$. In other words, there is no indication of either spin or valley degeneracy breaking. The LL degeneracy on the electron side of the neutrality point could not be determined due to an insufficient number of oscillations within the experimentally accessible carrier density and magnetic field ranges.

\section{Conclusions} 

For all hydrogenated samples, the $\nu = -2$ filling factor was observed and a four-fold spin and valley LL degeneracy at filling $|\nu|>2$ was observed in SdH (with the exception of samples HG70 and HG74, where insufficient number of SdH oscillations were observed). In other words a sequence $\nu = -2, -6,...$, leads us to conclude that the LL sequence on the hole side in hydrogenated graphene corresponds to that of pristine graphene with a topological Berry phase $\beta = \pi$. This phase, or equivalently the pseudo-spin winding number, is thus found to be robust in the presence of hydrogenation at an H/C ratio $\approx 0.07\pm0.02\%$. We emphasize that the robustness of the topological Berry phase is remarkable in two aspects. First, it survives unexpectedly for sufficient disorder to impart insulating behaviour in hydrogenated graphene. Second, it survives even if hydrogenation opens a gap by locally breaking sub-lattice symmetry. Although we do not have direct experimental evidence for this type of symmetry breaking in our samples, one cannot exclude this scenario in view of recent evidence for gap opening in ARPES measurements \cite{Grueneis10} and local sub-lattice symmetry breaking in STM experiments \cite{Grueneis12, Goler}. However, further experimental work is required to fully elucidate the relationship between adsorbate ordering, sub-lattice symmetry, energy gap and Diracness in 2D materials \cite{Diracness}.

\begin{acknowledgments}
We thank T. Pereg-Barnea, S. Roche and A. Champagne for useful discussion, and R. Talbot, R. Gagnon, G. Jones, J. Pucci and T.P. Murphy for outstanding technical support. This work was funded by NSERC, CIFAR, FRQNT, RQMP and the CRC program. A portion of this work was performed at the National High Magnetic Field Laboratory which is supported by NSF Cooperative Agreement No. DMR-0084173, the State of Florida, and the DOE, and we acknowledge the support of the LNCMI-CNRS in Toulouse, member of the European Magnetic Field Laboratory (EMFL). S. H. acknowledges support by the Italian Ministry of Foreign Affairs (Ministero degli Affari Esteri, Direzione Generale per la Promozione del Sistema Paese, progetto: Nanoelettronica quantistica per le tecnologie delle informazioni), and from the European Union Seventh Framework Programme under grant agreement no. 604391 Graphene Flagship. G.~G. acknowledges a short--term mobility grant from CNR.
\end{acknowledgments}

\end{document}